IAC-19-C1.3.8

# Accurate and Efficient Propagation of Satellite Orbits in the Terrestrial Gravity Field

**Elena Fantino[a*], Roberto Flores[b], Amna Adheem[c]**

[a] *Department of Aerospace Engineering, Khalifa University of Science and Technology, Abu Dhabi, United Arab Emirates, P.O. Box 127788*, elena.fantino@ku.ac.ae
[b] *International Center for Numerical Methods in Engineering (CIMNE), Building C1, Campus Norte UPC, Gran Capitán s/n, 08034 Barcelona, Spain*, rflores@cimne.upc.edu
[c] *Department of Industrial and Systems Engineering, Khalifa University of Science and Technology, Abu Dhabi, United Arab Emirates, P.O. Box 127788*, amna.adheem@ku.ac.ae
\* Corresponding Author

**Abstract**

Fast and precise propagation of satellite orbits is required for mission design, orbit determination in support of operations and payload data analysis. This demand must also comply with the different accuracy requirements set by a growing variety of scientific and service missions. This contribution proposes a method to improve the computational performance of orbit propagators through an efficient numerical integration that meets the accuracy requirements set by the specific application. This is achieved by (1) appropriately tuning the parameters of the numerical propagator (relative tolerance and maximum time step), (2) establishing a threshold for the perturbing accelerations (Earth's gravitational potential, atmospheric drag, solar radiation pressure, third-body perturbations, relativistic correction to gravity) below which they can be neglected without altering the quality of the results and (3) implementing an efficient and precise algorithm for the harmonic synthesis of the geopotential and its first-order gradient (i.e., the gravitational acceleration). In particular, when performing the harmonic synthesis, the number of spherical harmonics to retain (i.e., the expansion degree) is determined by the accuracy requirement. Given that higher-order harmonics decay rapidly with altitude, the expansion degree necessary to meet the target accuracy decreases with height. To improve the computational efficiency, the number of degrees to retain is determined dynamically while the trajectory is being computed. The optimum expansion degree for each altitude is determined by ensuring that the truncation error of the harmonic synthesis is below the threshold acceleration. The work is a generalization to arbitrary orbits of a previous study that focused on communication satellites in geosynchronous inclined orbits. The method is presented and a set of test cases is analysed and discussed.

**Keywords:** Orbit propagation; Perturbations; Spherical harmonics; Terrestrial gravity field; Accuracy; Efficiency

## 1. Introduction

Satellite trajectory predictions are necessary for targeting, guidance, and navigation. The expansion of the space sector with missions of growing complexity introduces increasingly strict performance requirements on tasks such as orbit propagation, determination and maneuvering. Space surveillance and tracking, which deal with the prediction of orbital motion for space debris objects at the most populated altitudes, adds further criticality to the aforementioned operations. The development and use of advanced tools to carry out these tasks with accuracy and efficiency is becoming mandatory.

Orbit propagation methods are divided into three categories: numerical, analytical, and semi-analytical. Numerical methods, also referred to as special perturbations, approximate the solution of the equations of motion. They are accurate but time consuming. Analytical propagation methods, or general perturbations, replace the original equations of motion with an analytical approximation that captures the essence of the motion over some limited time interval. Approximating the motion makes analytical integration possible, which can be performed much faster than numerical integration. The drawback is a lower accuracy. Eventually, semi-analytical propagation methods blend numerical and analytical approaches. Theory and formulations of the three categories of methods are presented in classical textbook such as [1, 2]. Nowadays, great effort is put in the development of efficient integration procedures, and considerable attention is devoted to finding the most suitable formulation to solve a specific problem (see, for instance [3] and references therein). Regarding accuracy, most literature focuses on long-term propagation and bases the sensitivity analysis on analytical approximations [4, 5].

Here, we present a technique for the accurate and efficient propagation of geocentric orbits with Cowell's method. The equations of motion are expressed in Cartesian coordinates in the J2000 Earth-centered equatorial frame and the numerical integrator used is a variable time-step Runge-Kutta scheme of seventh order.

The accuracy demanded by the given application (e.g., a requirement set by the mission) is met by including in the equations of motion only the perturbing accelerations that produce appreciable effects on the orbit over the propagation interval and by appropriately setting the parameters of the numerical integrator. In this way, the amount of computations is minimized, and efficiency is achieved. Special care is put in modeling the contributions of the harmonics of the geopotential. The maximum harmonic degree to be included in the gravitational acceleration is determined on the basis of the desired accuracy level. This is done dynamically, as computations progress, through an appropriate function of the altitude. Further performance improvements are implemented to solve the singularity at the poles appearing when the traditional formulation of the geopotential in Associate Legendre Functions (ALFs) is used.

This work is the development of previous investigations which focused on one category of orbits (geosynchronous and highly inclined, see [6]), and, hence, on a specific range of altitudes. Here, we generalize the technique and we make it applicable to any altitude and accuracy level.

The paper is structured as follows: Section 2 illustrates the physical model; Section 3 discusses the numerical setup of the propagator for the performance test of the method on a Molniya orbit. Discussion and conclusions can be found in Sect. 4.

**2. The physical model**

The physical model accounts for
- the acceleration $a_E$ due to the gravity field of the Earth;
- the accelerations $a_M$ and $a_S$ caused by the third-body perturbations of Moon and Sun, respectively;
- the perturbation $a_{SRP}$ due to the solar radiation pressure;
- the term $a_R$ associated to the relativistic correction to gravity.

The acceleration $a$ of the satellite is the sum of the above contributions, i.e.,:

$$a = a_E + a_M + a_S + a_{SRP} + a_R. \quad (1)$$

Since the effect of the atmospheric drag on the orbit is not accounted for in the model, the current version of the orbit propagator can predict the trajectory evolution to a minimum altitude of, say, 1000 km (where the atmospheric density can safely be neglected).

*1.1 Terrestrial gravitational acceleration*

The conventional representation of the gravitational potential $V$ at a point $P$ in outer space is based on a spherical harmonic expansion in ALFs [7]

$$V(r, \varphi, \lambda) = \frac{GM_E}{r} \sum_{n=0}^{N} \left(\frac{R_E}{r}\right)^n \sum_{m=0}^{n} P_n^m(\sin\varphi)(C_{nm}\cos m\lambda + S_{nm}\sin m\lambda), \quad (2)$$

where $r$, $\varphi$ and $\lambda$ are the Earth-centred, Earth-fixed spherical equatorial coordinates of $P$ (respectively, radial distance, latitude and longitude from the fundamental meridian), $R_E$ is the mean Earth's radius, $GM_E$ is the Earth's gravitational parameter (the product of the Universal gravitational constant $G$ and the mass of the Earth $M_E$), the quantities $C_{nm}$ and $S_{nm}$ are the Stokes coefficients and $P_n^m(sin\,\varphi)$ is the ALF of the first kind of degree $n$ and order $m$. In Eq. (2), the series is truncated at a maximum degree $N$, called expansion degree. Then, $a_E$ is computed as the gradient of the geopotential:

$$a_E = \nabla V. \quad (3)$$

In the representation with ALFs, the latitudinal derivative of $V$, $\partial V/\partial\varphi$, is singular when $\varphi = \pm 90°$, which causes loss of precision for near-polar orbits. Since the computational performance is the focus of the proposed method, Eq. (2) has been replaced with a representation in Cartesian Earth-centred, Earth-fixed coordinates based on Helmholtz polynomials $H_n^m$ [8, 9]. The method is due to [10] and is singularity-free. Additionally, the implementation incorporates improved recursion schemes on the $H_n^m$'s and handles the sums by accumulating so-called lumped coefficients, which are harmonic sums over the degree. Under certain circumstances, e.g., when performing simulations over a latitude-longitude grid of points, the latter yields increased performance. The expression for the geopotential $V(r,\varphi,\lambda)$ is

$$V(r, \varphi, \lambda) = \sum_{m=0}^{N}(A_m \cos m\lambda + B_m \sin m\lambda)\cos\varphi^m, \quad (4)$$

in which

$$A_m = \sum_{n=m}^{N} \rho^n C_{nm} H_n^m, \quad (5)$$

$$B_m = \sum_{n=m}^{N} \rho^n S_{nm} H_n^m, \quad (6)$$

whereas the parallactic factor $\rho^n$ is defined as

$$\rho^n = \left(\frac{GM_E}{r}\right)\left(\frac{R_E}{r}\right)^n. \quad (7)$$

Formulas for the derivatives of $V$ (i.e., the cartesian components of the accelerations) and more details on the computation of the polynomials can be found in [11].

With the issue of the singularity at the poles solved, the computation of quasi-polar or polar orbits can be carried out without loss of precision.

The adopted gravity model is the zero-tide version of the Earth Gravitational Model EGM2008 [12]. It consists in a table of fully-normalized, dimensionless Stokes coefficients and their error standard deviations. The model is complete to degree and order 2159. However, at any given altitude $h$ the synthesis of gravitational acceleration is affected the uncertainty in the determination of the model coefficients. The accuracy of the coefficients is characterized by their standard deviations $\sigma(C_{nm})$ and $\sigma(S_{nm})$. We have used these dispersions and the central limit theorem to compute the standard deviation of the three components of the acceleration as a function of $N$ at each point of a 36x17 spherical grid (10° resolution in both latitude and longitude) at a given altitude $h$. The maximum error at each altitude is a measure of the intrinsic accuracy of the model for a given expansion degree and height. For example, at an altitude $h = 2000$ km we obtained the model error shown by the red dashed line of Fig. 1. The continuous curves (green, purple and yellow) represent the truncation error in the three components of the acceleration, i.e., the error incurred by neglecting all harmonics of degree $N$ or higher. Clearly, attempting to reduce the truncation error below the intrinsic uncertainty of the model is a waste of computing resources. At each altitude, there exists a maximum meaningful expansion degree $N_{max}$. At $h = 2000$ km, $N = 42$ is the expansion degree for which the truncation error equals the intrinsic model accuracy. Therefore, $N_{max} = 42$ and lower error bound is $3.84 \cdot 10^{-10}$ m/s$^2$.

The procedure to estimate $N_{max}$ has been applied over a wide range of altitudes (from 250 to 64000 km). Figure 2 shows the standard deviation of the intrinsic acceleration error for the EGM2008 model as a function of height. Using a 95% confidence level, the lower bound of the acceleration error is two standard deviations. Next, we determined, for each altitude, the expansion order yielding the same truncation error. Figure 3 illustrates $N_{max}$ as a function of $h$.

*1.2 Third-body perturbations*

The gravitational attraction of the third body (in this case, Sun or Moon) produces an acceleration $a_{3B}$ which can be expressed as follows [13]:

$$a_{3B} = -GM_B \left( \frac{r_S - r_B}{|r_S - r_B|^3} + \frac{r_B}{r_B^3} \right), \quad (8)$$

where $M_B$ is the mass of the perturbing body, whereas $r_B$ and $r_s$ are the geocentric position of the perturbing body and the spacecraft, respectively.

Evaluation of Eq. (8) requires knowledge of the geocentric position of the Sun and the Moon. Since the forces that these two bodies exert on the spacecraft are much smaller than the attraction of the Earth, it is not necessary to determine their coordinates to the highest precision when calculating the perturbing acceleration acting on the satellite. Approximate positions accurate to about 0.1-1\% are sufficient.

Here, Eq. (8) is particularized for the case of the Moon and the Sun, yielding the corresponding accelerations $a_M$ and $a_S$. The geocentric position vector of the Sun is determined from the heliocentric position vector of the Earth-Moon barycenter. The reference frame is the mean ecliptic and equinox of J2000. The model is approximate, it propagates the Keplerian elements assuming fixed rates. Next, it transforms the orbital position (perifocal coordinates) to the heliocentric ecliptic reference frame and finally to equatorial coordinates. Details, data and formulas are available at the Solar System Dynamics website [14].

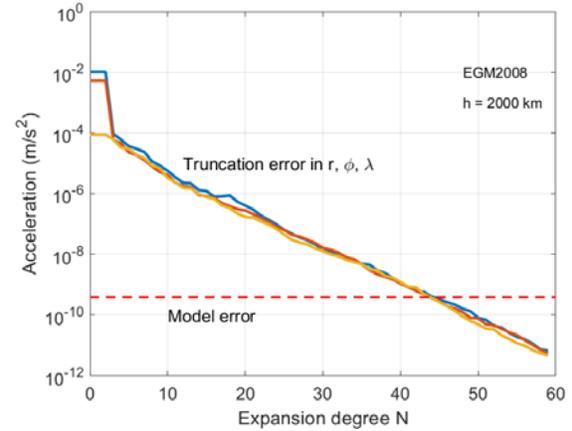

Fig. 1. Truncation and intrinsic model errors for EGM2008 at $h = 2000$ km.

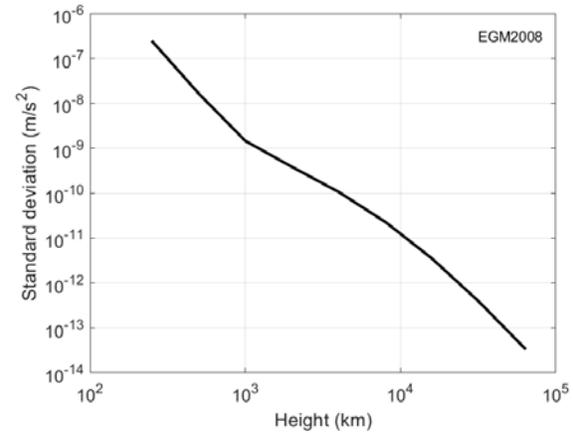

Fig. 2. Standard deviation of the acceleration error for the EGM2008 model as a function of altitude.

Simulation of the orbital motion of the Moon is carried out by assuming a set of mean orbital elements with respect to the mean ecliptic and equinox of J2000 and taking into account the linear regression of the ascending node and the linear precession of the line of apsides. This is followed by rotation to equatorial coordinates. The mean orbital elements and the rates of the two angular quantities are publicly available through the Solar System Dynamics website [15].

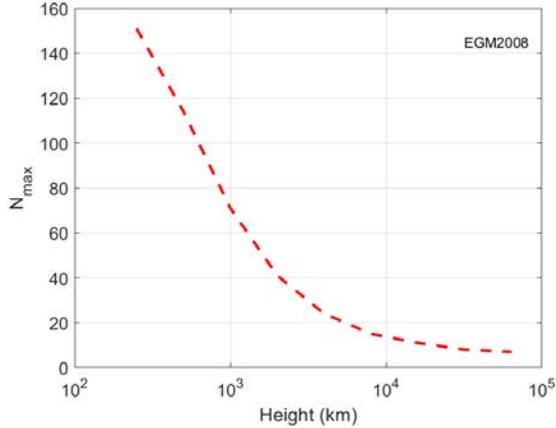

Fig. 3. Maximum meaningful expansion degree $N_{max}$ of EGM2008 vs height for a 95% confidence level.

*1.3 Solar radiation pressure*

As for the interactions of the spacecraft surface with the solar radiation, a spherical shape with rectangular wing-like solar panels has been assumed. When the surface normal is oriented in the direction of the Sun the following simplified formula for the acceleration $\boldsymbol{a}_{SRP}$ can be applied [13]:

$$\boldsymbol{a}_{SRP} = -f\left(\frac{A}{m}\right)\left(\frac{P_S}{4\pi d_S^2 c}\right)(1+k)\boldsymbol{u}_S. \tag{9}$$

$P_S = 3.846 \cdot 10^{26}$ W is the luminosity the Sun, $d_S$ is the Sun-spacecraft distance, $c$ is the speed of light in vacuum, $A/m$ is the front area-to-mass ratio of the satellite, $k$ is the surface reflectivity (ranging from 0 for complete absorption to 1 for specular reflection) and $\boldsymbol{u}_S$ is the spacecraft-to-Sun unit vector. The symbol $f$ represents the shadow factor, computed according to the double-cone model for solar eclipses [13]: $f = 0$ in umbra, $f = 1$ in sunlight, $0 < f < 1$ in penumbra.

*1.4 Relativistic effects*

The effects of General Relativity can be included by adding a perturbation $\delta V$ to the Newtonian gravitational potential with the form [16]:

$$\delta V = -\frac{GM_E H^2}{c^2 r_S^3}, \tag{10}$$

in which $H$ denotes the spacecraft's specific orbital angular momentum. Hence, the relativistic contribution to the acceleration is

$$\boldsymbol{a}_R = \frac{3GM_E H^2}{c^2 r_S^5}\boldsymbol{r}_S. \tag{11}$$

**3. Tests**

Tests of performance (accuracy and computing time) have been conducted on the propagation of one specific orbit.

*3.1 Orbit selection*

The selected orbit is of Molniya type [17]. It could be the orbit of a communications satellite providing service to high-latitude regions in the northern hemisphere (see Fig. 4). The perigee has been assigned a height $h_\pi$ of 1000 km altitude and the orbital period is equal to half a sidereal day ($T = 43082.05$ s). The initial orbital elements are:

- Semimajor axis $a = 26562.85$ km
- Eccentricity $e = 0.7222$
- Inclination $i = 63.4°$
- Right ascension of the ascending node $\Omega = 0°$
- Argument of the perigee $\omega = 270°$
- Epoch of pericenter passage $\tau =$ JD 2458757.5 (Oct 1st 2019, 0 UT)

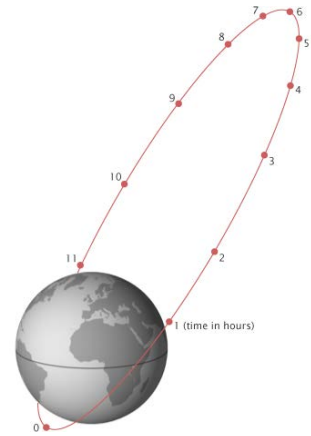

Fig. 4. Molniya orbit. The positions of the satellite at constant time intervals of 1 hour are marked.

The apogee altitude $h_a$ is 39367.43 km. For an orbit of this kind, station-keeping to within 1° accuracy over an interval of 1 month is sufficient for the intended applications. As a matter of fact, a satellite angular displacement of 1° corresponds to a shift in the ground track of some 100 km. This is a small distance compared to the size of the coverage area which typically extends over several thousand kilometers.

*3.2 Numerical setup*

In order to carry out an accurate and efficient orbit propagation, the parameters of the numerical integrator must be appropriately tuned and the relevant perturbing accelerations must be identified and included in the equations of motion.

The numerical integrator used in this work is a variable time step Runge-Kutta of seventh order. Its step-size limits $s_{min}$ and $s_{max}$ and the relative-error tolerance $\varepsilon$ are selected in such a way as to ensure that the position error $\Delta p$ accumulated over one orbital period $T$ of the initial osculating orbit is much smaller than the accuracy $\alpha$ required by the specific application, which is dictated by the orbit and the mission requirements. Both the step size $s$ and $\varepsilon$ must be small enough to guarantee the achievement of the target accuracy, but sufficiently large to minimize the number of steps and, as a result, the execution time. The values for $s_{min}$, $s_{max}$ and $\varepsilon$ are searched in a pre-processing phase through numerical experiments.

Also the choice of the perturbations to model depends on the accuracy level. In this case, the requirement on the position error per orbital period is used to define the threshold acceleration (the aforementioned $\hat{a}$) that the simulation must be able to sense, i.e., that the model must include. The value of $\hat{a}$ is determined by the conservative approach that a constant acceleration equal to $\hat{a}$ acting over $T$ causes a displacement equal to $\Delta p$.

*3.3 Accuracy requirement*

A uniform drift of 1° in position over 1 month is equivalent to 0.0167° per orbit. At the perigee altitude, this corresponds to a distance of 290 m per orbit. To detect reliably a deviation of this magnitude, the error of the trajectory propagation should be at least 10 times smaller (i.e., 30 m/orbit). Through numerical experiments, it has been determined that using a relative tolerance $\varepsilon$ of $10^{-6}$ and a maximum time step $s_{max}$ of 600 s produces errors well below the required 30 m/orbit. A constant acceleration of $10^{-8}$ m/s$^2$ acting for one orbital period produces a displacement smaller than 10 m. Therefore, perturbing accelerations under $10^{-8}$ m/s$^2$ can be discarded without altering the quality of the results. In fact, since most perturbations are likely to be cyclic instead of constant, their cumulative effect is expected to be even smaller.

The perturbation due to atmospheric drag can be neglected for this range of altitudes. The relativistic term varies between $10^{-11}$ and $10^{-8}$ m/s$^2$, respectively between apogee and perigee. Hence, it can be safely ignored too, being at most of the same magnitude as the admissible error. For standard communication satellites with an area-to-mass ratio of 0.01 m$^2$/kg, the effect of solar radiation pressure yields accelerations under $10^{-7}$ m/s$^2$, hence barely appreciable and with a marginal effect on the station-keeping requirements. Third-body effects are maximum at apogee, where the Sun contributes with accelerations at the level of $10^{-6}$ m/s$^2$ and the Moon gives effects almost one order of magnitude higher.

When performing the harmonic synthesis of the Earth's gravitational acceleration, the number of spherical harmonics to retain (i.e., the expansion degree) is determined by the accuracy requirement. Given that harmonics decay rapidly with altitude, the expansion degree necessary to meet the target accuracy decreases with height. To make the computations as efficient as possible, the number of degrees to retain is determined while the trajectory is being computed. To determine the optimum expansion degree at each altitude, a 10° by 10° grid is setup on a sphere of radius $R_E+h$. The expansion degree required to reduce the truncation error below a given threshold acceleration $\hat{a}$ at each point of the grid is defined as the degree for which the difference between the complete EGM2008 model and the truncated model is less than $\hat{a}$. The maximum among all the points of the grid is then determined and the procedure is repeated for several altitudes in geometric progression from 250 km up to 64000 km. For a given $\hat{a}$, this produces a table containing the required expansion degree as a function of altitude. The data are interpolated using an algebraic law (a combination of a power law and linear splines) which is then used by the numerical integrator to compute the gravitational acceleration at any altitude, guaranteeing the required accuracy while keeping the computational cost at a minimum. Figure 5 illustrates the expansion degree required to reduce the truncation error of the EGM2008 model below different levels (between $10^{-9}$ and $10^{-5}$ m/s$^2$) as a function of height. Note that $N$ should never be larger than $N_{max}$, so compatibility with the bounds stablished in Fig. 3 must verified.

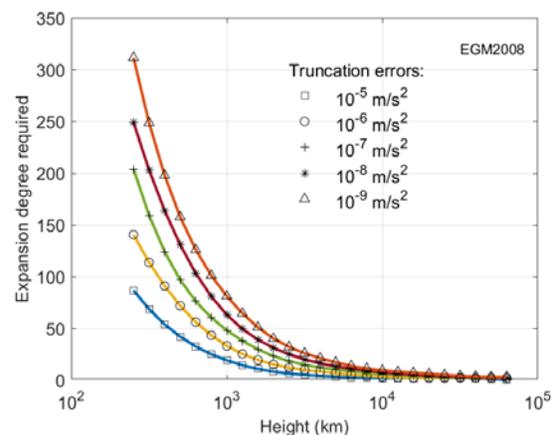

Fig. 5. Expansion degree required, as a function of altitude, to reduce the truncation error below a given threshold.

*3.4 Simulations*

The orbit chosen spans a wide range of altitudes. Therefore, the required expansion degree $N$ for the computation of the gravitational acceleration varies considerably over time. To make the effect of the degree of the harmonic synthesis easier to visualize, the numerical setup of the integrator has been changed slightly from the optimum values in Sect. 3.2. The maximum time step $s_{max}$ has been reduced from 600 s to 200 s, whereas the error tolerance $\varepsilon$ has been kept at $10^{-6}$. This improves the accuracy of the integrator to around 1 m per orbit, so small that it is negligible compared with the errors due to changes in the expansion degree. This makes the results easier to interpret. The propagation has been repeated for five different scenarios:

1. Dynamic $N=N(h)$ law,
2. Fixed $N=N(h_a)=3$,
3. Fixed $N=N(h_\pi)=64$,
4. Fixed $N=N_{max}(h_\pi)=71$
5. Fixed $N=100$.

The first scenario uses the degree-versus-altitude law corresponding to a truncation error of $10^{-8}$ m/s$^2$ (starred purple curve of Fig. 5). The second scenario uses the lowest expansion degree (the value corresponding to the apogee of the orbit), while the third run sets the maximum degree (i.e., the perigee value) at all times. The fourth case uses the maximum relevant expansion degree at the perigee altitude. This establishes the limit of the accuracy of the gravitational acceleration that can be achieved with the EGM2008 model. Finally, scenario 5 is a reference solution serving as baseline to determine the errors. It uses an expansion degree that is larger than the meaningful threshold (i.e., larger than $N_{max}$). Thus, for practical purposes, this case is free from truncation errors and can be used to estimate the accuracy of the other solutions. The remaining perturbations (solar, lunar, etc.) have been treated in the same way for all cases (according to the guidelines of Sect. 3.2). Simulations have been run in a current laptop processor (Core i7-7820HQ).

Scenario 1 is expected to offer the same accuracy as 3 (and much better than 2), but with a reduced computational cost. Case 4, on the other hand should yield virtually the same solution as 5, as there is no advantage in going over $N_{max}$. The results, reported in Table 1, confirm these expectations. The trajectories have been propagated for one month (i.e., 60 orbits) in order to improve the reliability of the CPU time measurement.

Table 1. Performance test results.

| Scenario | N | Error (m/month) | CPU time (s/month) |
|---|---|---|---|
| 1 | $N(h)$ | 120 | 0.672 |
| 2 | $N(h_a)=3$ | 6500 | 0.313 |
| 3 | $N(h_\pi)=64$ | 0.25 | 5.67 |
| 4 | $N_{max}(h_\pi)=71$ | 0.046 | 6.88 |
| 5 | 100 | - | 12.2 |

**4. Discussion and conclusions**

As shown in Table 1, the first and third runs meet the accuracy target. Both are well within the prescribed tolerance (30 m/orbit or 1800 m/month). It is noteworthy that the results of cases 4 and 5 are essentially identical. Including spherical harmonics above the meaningful expansion degree $N_{max}$ yields no improvement in the solution. The second scenario, on the other hand, sees the error increase by two orders of magnitude with respect to scenario 1, and does not match the accuracy requirements. Focusing on the computational cost of propagation, the variable-degree run requires almost one order of magnitude less CPU time than the third case, while retaining the same level of accuracy. In fact, the cost of the first scenario is comparable to the second (they differ just by a factor of two).

We demonstrated that it is possible to setup the numerical and physical model parameters in order to guarantee *a priori* a certain level of accuracy. Furthermore, the harmonic synthesis of the gravitational acceleration can be optimized through dynamic adjustment of the expansion degree. The procedure preserves the accuracy of the solution, while drastically reducing the computational cost in the case of highly eccentric orbits (where the optimum expansion degrees at perigee and apogee differ greatly). The methodology has been illustrated using an adaptive Runge-Kutta integrator, but it is quite general and applicable to any type of numerical propagator.

We also showed that there is an intrinsic limit in the accuracy of the gravitational acceleration that can be achieved with a given physical model (e.g., EGM02008). This limitation must be considered when assessing the suitability of the model for a particular application.

**Acknowledgements**

The work of E. Fantino and A. Adheem has been funded by Khalifa University of Science and Technology's internal grants FSU-2018-07 and CIRA-2018-85. The authors thank Prof. Jesús Peláez and Prof. Martín Lara for their suggestions.